\documentclass[aps,prl,twocolumn,superscriptaddress]{revtex4}

\begin{document}



\title{The structure of a single sharp quantum Hall edge probed by
momentum-resolved tunneling}


\author{M.~Huber}
\affiliation{Walter Schottky Institut, Technische Universit\"at
M\"unchen, D-85748 Garching, Germany\\}

\author{M.~Grayson}
\email{grayson@wsi.tu-muenchen.de}
\affiliation{Walter Schottky Institut, Technische Universit\"at
M\"unchen, D-85748 Garching, Germany\\}

\author{M.~Rother}
\affiliation{Walter Schottky Institut, Technische Universit\"at
M\"unchen, D-85748 Garching, Germany\\}

\author{W.~Biberacher}
\affiliation{Walther-Meissner-Institut, BADW M\"unchen, D-85748
Garching, Germany\\}

\author{W.~Wegscheider}
\affiliation{Universit\"at Regensburg, D-93040 Regensburg,
Germany\\}

\author{G.~Abstreiter}
\affiliation{Walter Schottky Institut, Technische Universit\"at
M\"unchen, D-85748 Garching, Germany\\}


\date{7 Nov 2004}


\begin{abstract}

Momentum resolved magneto-tunnelling spectroscopy is performed at a single
sharp quantum Hall edge.  We directly probe the structure of individual
integer quantum Hall (QH) edge modes, and find that an epitaxially overgrown
cleaved edge realizes the sharp edge limit, where the Chklovskii picture
relevant for soft etched or gated edges is no longer valid.  The Fermi
wavevector in the probe quantum well probes the real-space position of the
QH edge modes, and reveals inter-channel distances smaller than both the
magnetic length and the Bohr radius.  We quantitatively describe the
lineshape of principal conductance peaks and deduce an edge filling factor
from their position consistent with the bulk value.  We observe features in
the dispersion which are attributed to fluctuations in the ground energy of
the quantum Hall system.

\end{abstract}


\maketitle



Momentum-resolved tunneling has been used to measure both the dispersion of
electronic excitations, as well as their momentum-resolved density of states
or spectral functions.  Spectral functions of two-dimensional (2D)
\cite{Murphy95} and one-dimensional (1D) \cite{Kardynal97} systems have been
experimentally measured, and the dispersion relations of 1D systems have
shown evidence for spin-charge separation \cite{Auslaender02}. In the
quantum Hall regime, Kang, {\it et al.} \cite{Kang00} fabricated a pair of
coplanar 2D systems laterally coupled through a tunnel barrier, and were
able to map out the dispersion of the integer quantum Hall (QH) edge by
probing one QH system with another, qualitatively matching the expected
sharp-edge dispersion \cite{Ho94}.  Deviations in the data from a simple
dispersion picture have been cited as evidence of new interaction effects
\cite{Mitraetal}.  This Letter implements a different tunneling geometry
consisting of two {\em orthogonal} quantum wells in order to provide new
information about the sharp QH edge as well as its spectral function.  In
our geometry we can measure for the first time the real-space positions of
the edge states at a sharp edge, and we present a lineshape analysis which
fits the anticipated spectral functions in the system.  We demonstrate how
the Chklovskii screening picture does not apply at a sharp edge, but we have
instead experimentally realized the type of sharp edge first envisioned by
Halperin in his original work on edge states \cite{Halperin82}.  We observe
an additional step in the dispersion which we attribute to fluctuations in
the quantum Hall ground energy.  Although single QH edge tunneling was
investigated previously in the high field limit in a different kind of
device \cite{Changnco} the low field limit where momentum resolved
conductance resonances are evident had remained unexplored until now.

In our cleaved-edge overgrown structure \cite{Huber0204}, two orthogonal
GaAs quantum wells (QW) intersect in a T-junction with a tunnel barrier at
the intersection (Inset, Fig.~\ref{cond}). A magnetic field $B$
perpendicular to the first well $QW^{\perp}$ identifies this as the quantum
Hall effect system under study, while the second well functions as the probe
quantum well or $QW^{\parallel}$. The first well is $w^\perp =$ 150~\AA\
wide with an electron density of $n^{\perp} = 1.9 (1.7) \times 10^{11} {\rm
cm}^{-2}$ for Sample I (II), and a mobility of $\mu^{\perp} \simeq 2 \times
10^6 {\rm cm}^2/{\rm Vs}$.  The second orthogonal well is grown after
cleaving this sample in situ in the growth chamber in the (110) plane
\cite{Pfeiffer90} and overgrowing a $b=50$~\AA\ wide and 0.3~eV high
Al$_{0.33}$Ga$_{0.67}$As tunnel barrier.  This $w^{\parallel}=200$~\AA\
quantum well is then grown with a density of $n^{\parallel} \simeq 2.3
\times 10^{11}{\rm cm}^{-2}$ and an estimated mobility of $\mu^{\parallel}
\simeq 1 \times 10^5 {\rm cm}^2/{\rm Vs}$.  The tunnel junction is extended
typically $20~\rm{\mu m}$ along the cleaved edge. The $QW$'s are separately
contacted with ohmic indium contacts and the tunnel current $I$ is studied
under applied bias $V$ in a $^3\rm{He}$ cryostat at temperatures of
$360~\rm{mK}$.  No temperature dependence was observed up to 1 K.

Fig.~\ref{cond} shows the differential tunnel conductance $\frac{dI}{dV}$
measured using lock-in techniques while sweeping the $B$-field. At zero bias
we observe well-developed peaks in $\frac{dI}{dV}$ at certain values of the
magnetic field, and we resolve up to four of these peaks (denoted with
$n=0,1,2,3$ from right to left).  Their width and height above the
background is larger for those at higher $B$-field, with the peaks showing a
slightly asymmetric shape with a steeper slope at the high $B$ side. The
zero-bias conductance is strongly suppressed above 5~T.

We explain these observations with momentum conserved tunneling between the
2D probe contact and the edge states of the QH system.  To build up a
spatially intuitive picture, we will express all dispersions in terms of an
orbit center coordinate, $X$. Translational invariance in $y$ guarantees
that $k_y$ is conserved upon tunneling, or identically that $X=k_y l_0^2$
the orbit center coordinate in $x$ is conserved, where $l_0^2 = \hbar / eB$
is the squared magnetic length.  The probe contact dispersion is
$E^{\parallel}(k_y,k_z)= \frac{\hbar^2}{2m^*} (k_y^2+k_z^2)$, or in terms of
the orbit center coordinate, $E^{\parallel}(X,k_z)=\frac{\hbar^2}{2m^*}
[(\frac{X}{l_0^2})^2+k_z^2]$.  To find the corresponding expression for the
QH system we introduce the $y$-translationally invariant Landau gauge ${\bf
A}=xB{\bf \hat{y}}$ for the magnetic field ${\bf B}= B{\bf \hat{z}}$ with
$x=0$ in the center of the probe $QW^{\parallel}$.  The wave function can
then be separated into product form $\Psi_{n,X}(x,y)=
exp(-iy\frac{X}{l_0^2}) \psi_{n,X}(x)$, with the $x-$component
$\psi_{n,X}(x)$ obeying the Schroedinger equation:

\begin{eqnarray}
\left[\frac{p_x^2}{2m^*}+\frac{1}{2}m^*\omega_c^2(x-X)^2+
\Phi(x,V)\right]\psi_{n,X}(x) \nonumber\\
{}{}{}{}{}{}{}=E^{\perp}_n(X)\psi_{n,X}(x)
\label{Schroedinger}\end{eqnarray}

\noindent The dispersion curve $E^{\perp}_n(X)$ for the $n^{\rm th}$ Landau
band is calculated for an infinitely sharp step function edge potential at
the left wall of the barrier: $\Phi(x,V)=eV + \Delta$ for $x \le -(b +
\frac{w^\parallel}{2}); = \infty$ for $x > -(b + \frac{w^\parallel}{2})$.  
$V$ is the applied voltage and $\Delta$ is the ground energy difference
between the two systems.  The resulting dispersions are plotted in
Fig.~\ref{edge modes} for two different magnetic fields for $V = 0$.  We
note that the orbit center $X$ {\it can} be inside the barrier since the
orbit itself always remains outside as a skipping orbit.  We neglect spin
splitting, a point which will be discussed later in this paper, as well as
the $B$ dependence of the dispersion $E^{\parallel} (X,k_z)$, which can be
shown to be negligible \cite{inplane_B}.




The resonance condition is achieved when the outermost Fermi point in the
probe intersects the Landau band dispersion.  The real space position probed
by the Fermi point is its orbit center $X_{\rm F}(B) = \hbar k_{\rm
F}^\parallel/eB = k_{\rm F}^{\parallel}l_0^2$, one cyclotron radius to the
left of the probe QW center.  The orbit center of the $n^{\rm th}$ Landau
band at the Fermi energy $\xi_n$ is defined by the condition
$E^\perp_n(\xi_n)=E_{\rm F}$. When these two coincide, $\xi_n = X_{\rm F}$,
the resonance gives rise to the $n^{\rm th}$ experimentally observed
conductance peak. With the probe Fermi momentum $|k^{\parallel}_F| = 1.2
\times 10^{8}~{\rm m}^{-1}$, the distance of this orbit center from the
barrier can be determined $X^b(B)=|X_F(B)| - b - \frac{w}{2}$ (top axis in
Fig.~\ref{cond}) and is listed for each measured resonance in Table
\ref{edge}.

The exact condition of resonance depends on the band offset $\Delta$ between
the two systems.  Both dispersions are fully determined at a given $B$, but
differing QW ground energies and stray electric fields at the junction may
shift the ground energy of one 2D system relative to the other, or slightly
alter the density of QW$^\perp$ at the junction \cite{Changnco}.  To
accomodate such effects we empirically shift the calculated QH dispersion by
an energy $\Delta$ until it intersects with the Fermi point of the probe,
satisfying $E^\perp_n(\xi_n)  = E^\parallel(X_{\rm F}, k_z=0) = E_{\rm F}$
as in Fig. \ref{edge modes}.  The resulting offset of $\Delta = 2.2$ meV is
accounted for principally by the ground energy difference between the two
square wells $\Delta_{QW} = \frac{\hbar^2\pi^2}{2 m^*}(\frac{1}{w^{\parallel
2}} - \frac{1}{w^{\perp 2}}) = 1.4$ meV.  We attribute the remaining
difference to stray electric fields across the barrier.  The consistency of
this fit is our first confirmation that we are reasonably within the sharp
edge limit.

Due to the abruptness of the edge potential, the Chklovskii picture of
(in)compressible strips is not valid at this experimentally realized sharp
edge.  According to Chklovskii, {\it et al.} \cite{Chklovskii92},
alternating compressible and incompressible strips will form at the edge of
a QH system if the inter-edge spacing is much greater than both the minimum
screening distance (the Bohr radius $a_0 = \frac{4 \pi \epsilon_r \epsilon_0
\hbar^2} {m^*e^2}$ = 10 nm in GaAs) and the wavefunction width (the magnetic
length $l_0 = \sqrt{\hbar/eB}$ in Table \ref{edge}).  The wavefunctions
shown in the right of Fig. \ref{edge modes} each have nodes at the barrier,
with the $n = 1$ branch wavefunction having one additional node and
extending further to the left than the $n = 0$ branch. Due to the sharp
confinement, the wave functions of these two branches share their rightmost
node and therefore {\it completely overlap}, making the interedge spacing
{\it less} than $l_0$ and $a_0$ in violation of the Chklovskii criterion.  
Even in the case of a single edge mode (Fig. \ref{edge modes}, left) no
compressible strip is expected to form since the edge state is within $l_0$
and $a_0$ of the hard wall. Just as compressible strips are not allowed to
form at a sharp edge, the same length scale and screening arguments forbid
edge reconstruction, whereby Coulomb interactions would cause a strip of
charge to separate from the edge \cite{Chamon94}.

The fixing of the resonance condition determines the Fermi energy at the QH
edge, which can be expressed in terms of an edge filling factor.  Analogous
to Ref. \cite{Kang00}, our simplest estimate for $\nu_{\rm edge}$ assumes
broadened spin-degenerate Landau bands giving a flat density of states $g(E)
= \frac{2eB}{h} \frac{1}{\hbar \omega_c} = \frac{m*}{\pi \hbar^2}$.  The
edge filling factor is then

\begin{equation}
\label{nuedge}
\nu_{\rm edge} = 2 \frac {E_{\rm F,edge}}{\hbar \omega_c}
\end{equation}

\noindent Fig. \ref{edge modes} shows the cases for $n = 0$ and 1 peaks,
giving $\nu_{\rm edge}$ estimates 2.0 and 3.8, respectively, in fair
agreement with the bulk filling factor at these fields, $\nu = 2.3$ and 4.1
(see also Table \ref{edge}).  Later we will see evidence that the density of
states $g(E)$ is not so flat as this simplified picture, yet the above
estimate offers another consistency check of the sharp edge picture.


We can also explain the lineshape of the prominent zero-bias conductance
peaks. The tunnel conductance at $V=0$ is proportional to the number of
states at the Fermi energy that overlap in momentum space.

\begin{equation}\label{spectral_integral}
\frac{dI}{dV}\sim
\sum_{k_y,k_y'}|t_{k_y,k_y'}(B)|^2A_{QHE}^{\perp}(k_y,E_F)A_{probe}
^{\parallel}(k_y',E_F)
\end{equation}

\noindent Solving the Schr\"odinger equation for the transmission and
assuming perfect momentum conservation, we find that the transmission
probability $|t(B)|^2 \sim B~\delta_{k_y,k_y'}$ is roughly proportional to
the magnetic field for low fields ($l_0>b$), contributing to the smaller
peak height observed at lower $B$.  To quantify the lineshape we first
assume $A_{QHE} ^{\perp}(k_y,E_F)\sim\sum_n \delta (E^{\perp}_F-E_n(k_y))$
for the spectral function of the quantum Hall system. In the probe contact
the component of the Fermi circle in the $k_z'$ direction results in a van
Hove-like singularity in the probe spectral function at the Fermi point
$k_y'=-k_F$: $A_{probe}^{\parallel} (k_y',E_F)~\sim~1/\sqrt{E_F-
\hbar^2k_y'^2/2m^*}$. The resulting calculated $dI/dV$ reproduces the
asymmetry observed in the experimental conductance of Fig.~\ref{cond} ({\it
dashed line}). As in Ref.~\cite{Eisenstein91} we can replace the
$\delta$-function in $t_{k_y,k_y'}(B)$ with a gaussian of full width $\Delta
k = 1.6 \times 10^{7} {\rm m}^{-1}$ to account for small
non-$k$-conservation, giving an excellent fit to the experimental data ({\it
dotted line}).  Differentiating the probe orbit center equation yields
$\Delta B = B \frac{\delta k}{k_{\rm F}}$, meaning the linewidth of the
resonances narrows with decreasing $B$ as observed.

The structure of the edge near zero bias can also be examined by sweeping
the $V$ for a series of fixed $B$ (Fig.~\ref{cond}).  A tunnel bias $V$
shifts the dispersion curves in energy with respect to each other, and from
Fig.  \ref{edge modes} one can see that an increasing negative bias
\cite{posbias} raises the Fermi point of the probe past the successive
Landau branches, with corresponding conductance peaks labelled $n = 0,1,2,3$
in Fig.~\ref{cond}.  Within a given Landau branch, increasing $B$ shifts the
resonance more towards negative $V$, allowing one to map out the entire
Landau band dispersion.  This behavior is explained in further detail in
Ref.~\cite{Huber0204}. Fig.~\ref{splitting} shows such a scan of $dI/dV$
where the peak positions in $V$ vs. $B$ map out the low-energy dispersion of
the edge modes in $E$ vs. $k$ with the maxima indexed as in Fig.~\ref{cond}.  
Most notable is the step in the $n=0$ dispersion curve which does not cross
the zero bias line continuously, but instead at $B=3.5$~T shows a splitting
of $\Delta V\sim 4~$mV.

We explain this feature with a fluctuating ground energy in the quantum Hall
edge. As shown in the inset of Fig. \ref{splitting}, the band diagram
corresponding to the peak condition changes discontinuously as the Fermi
energy near the edge jumps between Landau levels $n=0$ and $1$ for $\nu_{\rm
edge} <2$ and $\nu_{\rm edge}>2$ respectively.  Upon increasing $B$ such
that $\nu_{\rm edge} < 2$, the peak condition requires an additional voltage
$\Delta V=-\hbar \omega_c/e = -6~{\rm meV}$ at 3.5 T
\cite{thermal_activation}.  The observed jump of -4 meV can be attributed to
disorder broadening which narrows the mobility gap in the density-of-states.  
Whereas in standard soft QH edges the compressible strips screen any bulk
ground energy oscillations from reaching the outermost edge, the observation
of a step at $n = 0$ demonstrates that sharp edges are unable to screen.  
Note that the measured dispersion curves for $n = 1,2,3$ show no step at $B
= 3.5~$T because at larger negative bias the edge depletes and the resulting
smooth edge potential gives rise to compressible strips that {\it do} screen
the bulk oscillations.  A second important observation is that the $\nu_{\rm
edge} = 2$ jump occurs at the edge at a $B$-field where we also expect a
$\nu = 2$ jump in the bulk. This is a second indication that the edge
filling factor $\nu_{\rm edge}$ is close to the bulk value $\nu$.

The additional shoulder at $B = 4.1$~T on the high $B$-field side of the
$n=0$ peak can not be explained within this model.  Translated into the
orbit-center coordinate $X$ the shoulder is separated from the main peak by
$\sim 3$~nm. Recalling the length scale comparisons above, this short
distance rules out that it is a signature of either standard edge
reconstruction or the Chklovskii (in)compressible strip picture. Instead of
resulting from real-space structure, it may result from structure in the
energy spectrum.  For example, it could possibly be an artifact of the
previously described $\hbar \omega_c$ jump in the chemical potential near
$\nu = 2$ ($B=3.5$~T), or it may be a signature of the exchange enhanced
spin-split gap \cite{Dempsey}.

In conclusion we have probed the QH edge state structure at a sharp cleaved
and overgrown edge.  We have directly measured the real space position of
the edge channel orbit centers and demonstrated that the Chklovskii picture
is not valid in this system. The prominent lineshape is fully described with
the spectral functions in the tunnel contacts if we include a gaussian
broadening in the momentum selection rule.  An edge filling factor is
deduced from conductance peak positions in $B$ and agrees with the bulk
value, implying uniform electron density up to the edge in these structures.  
Evidence for a jump in the chemical potential confirms that $\nu_{\rm edge}
= \nu$, and that this sharp edge cannot screen bulk electrostatics.  The
existence of chemical potential oscillations may be important for
interpreting the peak lineshapes in double-edge tunneling geometries
\cite{Kang00}, and the extension of the bulk filling factor all the way to
the sharp edge has important implications on previous tunnel experiments on
different cleaved quantum Hall edge structures \cite{Changnco}.

The characterization of low B-field momentum resolved tunneling in this
device opens the door for future proposed experiments \cite{Zulicke02} to
measure fractional QH correlations at high magnetic fields in this device,
with the advantage over other experiments that the edge filling factor can
be determined using the methods explained here.

M. H. and M. G. gratefully thank U. Z\"{u}licke and M. Geller for
stimulating discussions. The work was supported by the German Physical
Society (DFG) via Schwerpunktprogramm Quanten-Hall-Systeme and the
EU-Research Training Network COLLECT.

\begin{table}[h]
\caption{
\label{edge}
(Sample I) Bulk filling factor $\nu$, edge filling factor 
$\nu_{\rm edge}$, as well as measured distance $X^b$ of the orbit
center from the tunnel barrier for each conductance resonance, 
compared to the magnetic length $l_0$.}
\begin{tabular}{lcccc}
$n$         	      &    0    & 1       &     2    &   3 \\
\hline
$\nu$       	      & $2.3$   & $4.1$   & $5.9$    & $7.7$ \\
$\nu_{\rm edge}$      & $2.0$   & $3.8$   & $5.7$    & $7.6$ \\
$B$         	  (T) & $3.44$  & $1.90$  & $1.33$   & $1.02$ \\
$X^b$   	 (nm) & $8\pm 2$&$26\pm 3$& $44\pm 5$&$62\pm 6$ \\
$l_0$     	 (nm) & $14$    & $19$    & $22$     & $25$ \\
\end{tabular}
\end{table}

\begin{figure}[h]
\caption{
\label{cond}
The differential tunnel conductance $dI/dV$ versus magnetic field $B$
at zero DC bias voltage for Sample I. The dotted and dashed curves
indicate the expected resonance lineshape with and without disorder
broadening. Inset: two quantum wells are arranged in a T-shape
separated by a $b=$50~\AA\ thick tunnel barrier; a magnetic field $B$
creates quantum Hall edges in $QW^{\perp} (w^\perp = 150$ \AA) probed
by tunneling from $QW^{\parallel} (w^\parallel = 200$ \AA).}
\end{figure}

\begin{figure}[h] 
\caption{
\label{edge modes} 
Calculated dispersions $E^{\perp}_n(X)$ and $E^{\parallel}(X,k_z)$ at
the magnetic fields of the experimentally observed $n=0$ (left) and
$n=1$ (right) zero-bias conductance peaks. The conductance peaks arise
from the resonance condition where the Fermi point in the probe
intersects with the Landau dispersion.  The occupied part of each
Landau band is in black, with the unoccupied part in grey.  The
conduction band potential is shown as a shaded grey background.  The
wavefunctions of each mode at $E_{\rm F}$ are depicted with thin solid
lines above the Fermi energy.} 
\end{figure}

\begin{figure}[h]
\caption{ 
\label{splitting}
$\frac{dI}{dV}$ vs. $V$ at different $B$ (0.05 T steps offset by 0.5
$\mu$S.) The conductance peak for the outermost $n=0$ edge channel splits
into two near 3.5~T, evidence that the jump in the bulk chemical potential
at $\nu = 2$ is being seen at the edge.  Peaks for the inner depleted
$n=1,2,3$ channels remain continuous at 3.5~T since the smoothly depleted
edges can screen the bulk $\nu = 2$ jump. (Sample II)}
\end{figure}


\begin{references}
\bibitem{inplane_B}
With an in-plane $B$-field of 3.5~T in a $w^{\parallel} =  200$~\AA\ 
wide quantum well the intersubband mixing is small enough to keep 
the distortion of $v_F$ and $k_F$ below 5 percent.
\bibitem{posbias}
Positive bias corresponds to tunneling from the QH edge to
the probe and will be addressed in a future work.
\end{references}
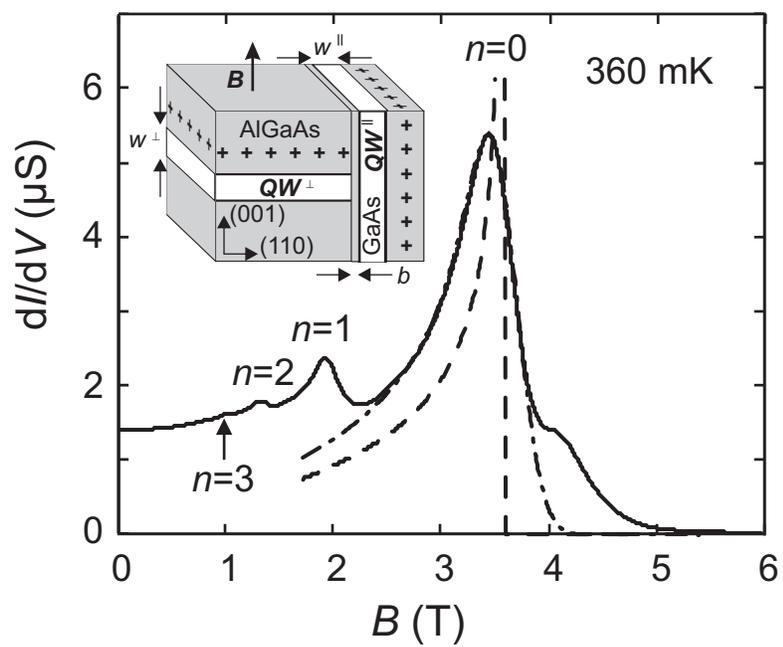
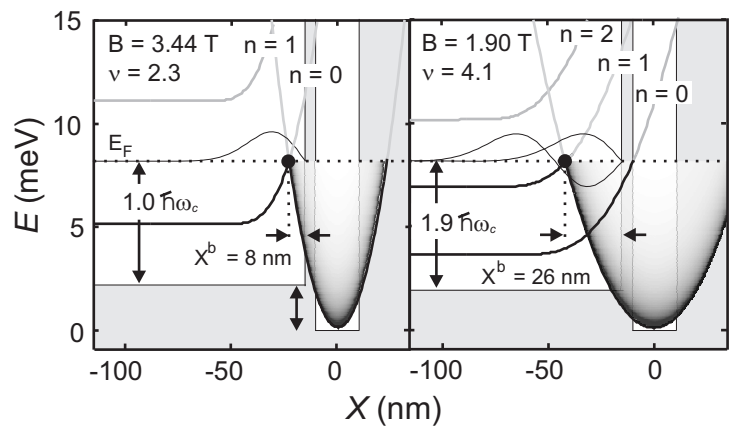
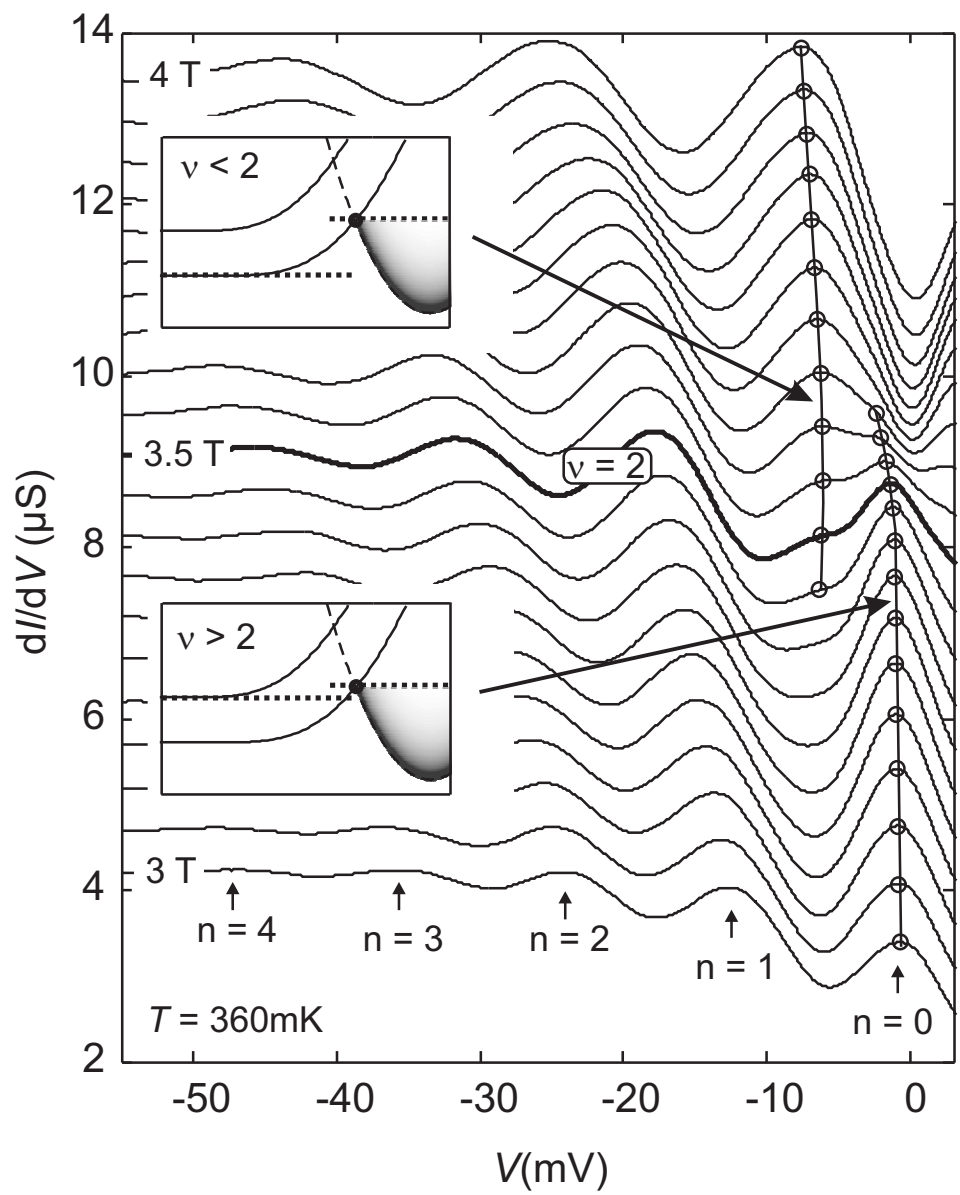
\end{document}